\documentstyle[preprint,prl,aps]{revtex}
\begin{document}
\preprint{\today}
\draft
%
%
\title{Testing temporal Bell inequalities through repeated measurements in
rf-SQUIDs}
\author{Tommaso Calarco${}^{1}$ and Roberto Onofrio${}^{2}$}
\address{${}^1$Dipartimento di Fisica, Universit\`a di Ferrara, and
INFN, Sezione di Ferrara, \\
Via Paradiso 12, Ferrara, Italy 44100\\
${}^2$Dipartimento di Fisica ``G. Galilei'',
Universit\`a di Padova and INFN, Sezione di Padova, \\
Via Marzolo 8, Padova, Italy 35131}
\date{\today}
\maketitle
%
%
\begin{abstract}
Temporal Bell-like inequalities are derived taking into account the influence
of the measurement apparatus on the observed magnetic flux in a rf-SQUID.
Quantum measurement theory is shown to predict violations
of these inequalities only when the flux states
corresponding to opposite current senses are not distinguishable.
Thus rf-SQUIDs cannot help to discriminate realism and quantum mechanics
at the macroscopic level.
\end{abstract}

%
%
%
%
When quantum mechanics is extended to the macroscopic world some
contradictions with realism, {\em i.e.} the prejudice according to which
objects exist regardless of their observation, are evident.
A deeper understanding of this contrast has relevance both to better
study quantum phenomena already occurring in the macroworld,
such as macroscopic quantum transport of particles in superfluidity
and superconductivity, and to understand the relationships among
quantum mechanics, macroscopic realism and classical physics, this last
being contained in the former but at the same time playing
a crucial role for the existence of the measuring apparatus.
It became evident that the relevant features under debate were testable
with numerical predictions and actual experiments \cite{BELL}.
For instance, spatial Bell inequalities have been tested
and the experimental results agreed
with the violation of the inequalities predicted by quantum mechanics
\cite{ASPECT}.
Although the interpretation of these results is still
under debate \cite{SANTOS}, the attention has been shifted in
recent years to test temporal Bell inequalities \cite{LEGG1}.
In this case the crucial difference is that a unique system undergoes
to repeated measurements at different times, unlike the case of spatial
Bell inequalities where two systems are subjected to unique and
simultaneous measurements.
Furthermore, the aim of temporal Bell inequalities, in the original spirit
of Leggett and Garg \cite{LEGG1}, was to test quantum mechanics at the
macroscopic
level whenever a macroscopic observable of the system is monitored.
This allows one to study the extension of quantum theory
to the macroscopic world to solve its paradoxical contrast
with the widely accepted realistic view \cite{LEGG2,LEGG3}.
Following this proposal, Tesche discussed in detail a concrete
experimental scheme based upon use of superconducting quantum
interferometer devices (SQUIDs) \cite{TE0,TE}.
The feasibility of any experiment aimed at testing macroscopic realism
through temporal Bell inequalities has been criticized due to the role
played by the concept of non-invasive measurements \cite{BA1,PE}.
In this letter we consider Bell inequalities for a measurement
of magnetic flux on a rf-SQUID at certain set of times
and the predictions of quantum theory including the effect of the previous
measurements in the evolution of the system.
We also consider the quantum limitations dictated by the uncertainty
principle to the measurement
of magnetic flux in the same set of measurements.
The two investigations are finally merged together to estabilish
if theoretically {\em predicted} violations of
temporal Bell inequalities can actually
{\em be observed} when the effect of the measurement is taken into account.

The system we are considering is an rf-SQUID where the magnetic
flux $\phi$ evolves in a bistable potential.
The corresponding Hamiltonian for the magnetic flux $\phi$
(in the unit system in which $\hbar=2m=1$, $m$ being the effective mass
of the system) is:
\begin{equation}
\label{HAM}
H= -\frac{\partial^2}{\partial\phi^2}
- {\mu \over 2} \phi ^2 + {\lambda \over 4} \phi ^4
\end{equation}
where $\mu$ and $\lambda$ ($\mu, \lambda >0$) are parameters associated
to the superconducting circuit.
The potential corresponding to the last two terms in (\ref{HAM})
has the shape of a double
well with minima at $\pm\Phi_{min}=\pm {(\mu/ \lambda)}^{1/2}$,
separated by a distance $\Delta L\equiv 2\Phi_{min}$.
The effective potential in (\ref{HAM}) can be rewritten in terms of
the minima and the energy barrier $|V(\Phi_{min})|=\mu^2/4\lambda$ as
\begin{equation}
\label{VEFF}
V(\phi)= 2 V(\Phi_{min})
\left[ 1-{1\over 2} \left({\phi\over \Phi_{min}}\right)^2\right]
\left({\phi \over \Phi_{min}}\right)^2.
\end{equation}
Both the distance between the two minima $\Delta L$ and the energy barrier
$|V(\Phi_{min})|$ depend upon the parameters $\mu$ and $\lambda$.
The whole analysis is carried out in a dissipationless
environment, in which quantum coherence can be observed.
Following Leggett and Garg \cite{LEGG1} we subdivide the values of
magnetic flux in the two regions $\phi>0$, $\phi<0$, respectively
corresponding to clockwise and counterclockwise senses
for the superconducting currents.
The probability for the observed magnetic flux $\Phi$ to correspond
to one definite sense of circulation for the current, for instance $\Phi>0$,
is defined as
\begin{equation}
P\{\Phi(t)>0\}={{{\int_0^{+\infty}} d\phi |\psi(\phi,t)|^2}\over
{{\int_{-\infty}^{+\infty}} d\phi |\psi(\phi,t)|^2}}
\end{equation}
where $\psi(\phi,t)$ is the time-dependent wavefunction of the
superconducting current in the magnetic flux representation.
It is possible to write also correlation probabilities for the results
of two measurements performed at times $t_i$ and $t_j$, with $t_{ij}=t_i-
t_j$ called quiescent time (we consider the
limit of impulsive measurements, having therefore a negligible duration,
situation well approximated in practice by fast switching superconducting
circuits), for instance
\begin{equation}
\label{PROBCORR}
P^{ij}_{+-}\stackrel{\rm def}{\equiv}P\{\Phi(t_i)>0, \Phi(t_j)<0\}.
\end{equation}
In a realistic model, in which the sign of the flux is defined even when not
measured, we can write temporal Bell-type inequalities such as
\begin{equation}
P^{bc}_{+-}\leq P^{ab}_{++}+P^{ac}_{--}
\label{BELLINEQ}
\end{equation}
where different histories for the possible measurements have been considered:
the magnetic flux not measured at $t_a$ and measured respectively
with positive and negative values at $t_b$ and $t_c$, flux measured with both
positive
values at $t_a$ and $t_b$ and not measured at $t_c$, flux measured at $t_a$
and $t_c$ with both negative values and not measured at $t_b$ (see Fig.~1).
Eq. (\ref{BELLINEQ}) can be rewritten in an alternative form, which
shows the dependence on the quiescent times:
\begin{equation}
\Delta P(t_{ab},t_{bc})=P^{bc}_{+-}-P^{ab}_{++}-P^{ac}_{--}
\leq 0.
\label{DELTAP}
\end{equation}
We want to check whether quantum mechanics predicts violations of eq.
(\ref{BELLINEQ}),{\em i.e.} if exists at least a pair of quiescent times
for which $\Delta P(t_{ab},t_{bc})>0$.

The effect of the measurement process is introduced by means of a non-unitary
filtering weight which selects a particular result of the measurement
with a given accuracy.
In this way the wavefunction at the end of an impulsive measurement
$\psi(\phi,t^+)$ is given by the wavefunction immediately before the
measurement $\psi(\phi,t^-)$ multiplied by a weight function $w_{\Phi}(\phi)$.
The square modulus of the output wavefunction  $\psi(\phi,t^+)$
is the probability of finding the system in the state given by $w_{\Phi}(\phi)$
itself. Following von Neumann \cite{vN} we write such a weight as
\begin{equation}
\label{WVN}
w_{\Phi}^{\it v.N.}(\phi) \propto \left\{
\begin{array}{lc}
1~~~ & {\rm if} \, |\phi-\Phi|<\Delta\Phi,\\
0~~~ & {\rm otherwise}
\end{array}
\right.
\end{equation}
where $2\Delta\Phi$, the width of the filter of the
meter, is hereafter called instrumental error.
Other choices for the filtering weight are possible.
For instance a less discontinuous, and
therefore more physical, weight function is written, as in \cite{ME}:
\begin{equation}
w_{\Phi}(\phi) \propto \exp
\biggl \{-\frac{(\phi -\Phi)^2} {2 \Delta \Phi^2} \biggr \}
\label{WA}
\end{equation}
where $\Delta \Phi^2$ assumes the meaning of a variance.
Also, a filter complementary to (\ref{WVN}),
which would leave unchanged the state only
if the magnetic flux is localized around
$\Phi$, is the analytical counterpart of the so called
null-result measurement scheme proposed in \cite{TE}.
In either choices a particular outcome is privileged with respect to the
other possible ones and this is reflected in the dynamical evolution
of the magnetic flux. Moreover, the unitary evolution
is broken during the measurement,
as one expects for a selective measurement in which one get rid of
all the possible alternatives incompatible with the measurement result.
The actual value of the proportionality constants in eqs. (\ref{WVN}) and
(\ref{WA}) does not matter, because the only relevant quantities in the
subsequent calculations are normalized probabilities. For instance the quantity
\begin{equation}
\label{pfi}
P(\Phi) =\frac{\parallel\psi_\Phi(t^+)\parallel^2}{
\int \parallel\psi_{\Phi^{\prime}}(t^+)\parallel^2 d\Phi^{\prime}}=
\frac{\parallel\psi_\Phi(t^+)\parallel^2}{
\int\!\!\int e^{-\frac{(\phi-\Phi^{\prime})^2}{\Delta \Phi^2}}
|\psi(\phi,t^-)|^2d\phi\, d\Phi^{\prime}}
=\frac{1}{\sqrt{\pi}{\Delta \Phi}}\parallel\psi_\Phi(t^+)\parallel^2
\end{equation}
represents the probability that the observed value of the magnetic flux is
$\Phi$, with an instrumental error $\Delta\Phi$, in the case of a Gaussian
weight function such as (\ref{WA}). It is also clear that, to
distinguish the two signs of the magnetic flux required to have a dichotomic
variable useful for building Bell inequalities, one has to work
with instrumental errors $\Delta \Phi$ less than the distance
between the two wells $\Delta L$. We will consider in the following
a system with fixed parameters $\mu$ and $\lambda$,
and therefore constant $\Delta L$, and variable instrumental
error $\Delta \Phi$.
This is equivalent to consider the opposite situation
of a constant instrumental error and variables parameters of the rf-SQUID,
since the relative magnitude between $\Delta \Phi$ and $\Delta L$ rules
the distinguishability issue in a single measurement.

If more measurements are performed the back-action of the previous ones has
to be taken into account and the distinguishability of the two
signs of the magnetic flux depends, besides the instrumental error,
upon the time intervals between consecutive measurements.
Suppose that the system is initially in a pure state described by
the wavefunction $\psi(\phi,0)$.
Let us assume that a series of $N$ measurements at $t_n\equiv nT$
($n=0,1,\ldots,N-1$), has been performed with fixed instrumental error
$\Delta\Phi$ and known results $\{\Phi_n\}$. Finally we
suppose to perform another
measurement at $t_N\equiv NT$. According to the (\ref{pfi}), the probability
for obtaining a result $\Phi_N$ in this last measurement is
\begin{equation}
\label{P}
P_{\left\{\Phi_n\right\}_{n \leq N-1}}(\Phi_N)=
\frac{1}{\sqrt{\pi}{\Delta \Phi}}\parallel
\psi_{\left\{\Phi_n\right\}_{n \leq N}}(t_N^+)\parallel^2,
\end{equation}
{\em i.e.} it is proportional to the squared norm of the wavefunction after the
$N^{\rm th}$ measurement. The analytical expression of this last is
\cite{CALARCO}
\begin{equation}
\label{PSIN}
\psi_{\{\Phi_n\}_{n\leq N}}(\phi,t_N^+) = \sum_{l,m,n_1,\ldots,n_N=1}^\infty
W_{mn_1}^{\Phi_N}W^{\Phi_{N-1}}_{n_1n_2}\cdots
W_{n_Nl}^{\Phi_{min}}\exp\left\{-\frac{i\Delta T}{\hbar}
\sum_{i=1}^{N}E_{n_i}\right\}c_l\,u_m(\phi)
\end{equation}
where the $E_i$, $u_i$ are respectively the energy eigenvalues and
eigenstates of the system, the $W_{ij}^\Phi(\Delta\Phi)$'s are the
matrix elements of $w_\Phi(\Phi)$ between energy eigenstates (expressed through
(\ref{WVN}) or (\ref{WA}) in terms of the instrumental error $\Delta \Phi$)
on the latter and the $c_l$'s are the
projections on them of the initial state $\psi(\phi,0)$.
All the relevant quantities depend upon $\Delta\Phi$ through
$W_{ij}^\Phi(\Delta\Phi)$ in eq.~(\ref{PSIN}).
If the effect of the measurement is taken into account in this way an effective
magnetic flux uncertainty, with respect to the result $\tilde{\Phi}$,
arises \cite{MEOP2}
\begin{equation}
\Delta \Phi_{\it eff}(\{\Phi_n\}_{n\leq N-1},N)^2=
2\int_{-\infty}^{+\infty}(\Phi_N-\tilde{\Phi})^2
P_{\{\Phi_n\}_{n\leq N-1}}(\Phi_N)d\Phi_N.
\label{DAEFFTH}
\end{equation}
The effective magnetic flux uncertainty takes into account,
 besides the instrumental error $\Delta\Phi$,
the back-action effect of the previous measurements.
For stroboscopic measurements with constant result, the effective uncertainty
$\Delta \Phi_{\it eff}$ tends to reach
an asymptotic value $\Delta\Phi_{\it eff}^{\rm as}$ which
is greater than the instrumental error $\Delta \Phi$,
due to the effect of the back-action of the meter on the measured system,
unless the system is monitored in a regime unaffected by the quantum
noise, {\em i.e.} when $\Delta \Phi \gg \sigma$ where $\sigma$ is the width of
the initial wavefunction $\psi(\phi,0)$,  or in a quantum nondemolition
way \cite{CAVES,BRAG}.
We have already identified the quiescent times $T$ for which repeated
measurements of flux are quasi-quantum nondemolition
ones \cite{CALARCO} as the multiples
of the tunneling period $T=2\pi\hbar/(E_2-E_1)$.
This is the reason why we have chosen $T$ as the quiescent time for the
preparatory sequence referred to in Fig.~\ref{fig1}.
The correlation probabilities (\ref{PROBCORR}) have been evaluated
by applying (\ref{P}), and choosing the parameters of the
potential in (\ref{HAM}) as $\mu=9.6$ and $\lambda=1.536$ (always in
the unit system in which $\hbar=1$), such that $\Phi_{min}=2.5$ and thus
$\Delta L=5$. The choice of the initial state $\psi(\phi,0)$ is unessential
because, after the optimal preparatory measurement sequence, the state
collapses around the measurement result, as discussed in \cite{MEOP2}.
Now we can calculate the {\em quantum} predictions for $\Delta P$
 using (3-6).
In Fig.~2 a comparison between the results obtained for the temporal
Bell inequality and the already-known spatial
Bell inequality \cite{BELL} is shown to be very similar in the
dependence upon the relevant parameters, the quiescent times
for the temporal case and the polarimeter angles for the
spatial case.

An analogous dependence upon the measurement time (expressed in units of
the tunneling period $T$) is shown in Fig. 3 for the effective
magnetic flux uncertainties associated to each of the three sequences
of measurement. The optimality is linked to the multiples of $T$: thus the
different combinations of measurements are correlated to
different  orientations of the optimal regions in the $(t_{ab},t_{bc})$
plane. For instance, in the case of sequence III of Fig. 1 (lowest plot
in Fig. 3), there lie along diagonal lines, corresponding to
$t_{ab}+t_{bc}$ multiple of the optimal periodicity $T$.

The exclusion among the regions of violation to Bell
inequalities  and the regions of distinguishability of the magnetic flux
is emphasized in Fig.~4 which is a synthesis of all our discussion.
Contour plots for the Bell inequality violation region, and for the
regions of distinguishability of left and right part of the barrier
for the sequences of Fig.~\ref{fig1},
are simultaneously shown in a $t_{ab}$-$t_{bc}$ plot.
The shaded areas indicates the pairs of quiescent times for which
$\Delta P(t_{ab},t_{bc})$ is greater than zero, {\em i.e.} Bell
inequalities are violated. The quasi-triangular
regions correspond to the set of couples of quiescent times for which the two
wells are resolved even after the measurements, {\em i.e.} all the three
effective uncertainties $\Delta\Phi_{+-}^{bc}$, $\Delta\Phi_{++}^{ab}$ and
$\Delta\Phi_{--}^{ac}$ are less than $\Delta L$.
No intersection among the various contours plots exists, {\em i.e.} for
the sequences of measurements for which quantum mechanics gives
predictions in contrast with that of a realistic theory, one
cannot simply speak about distinct states because the effective uncertainty
does not allow one to distinguish them.
This result has been tested with respect to a certain number of conditions.
Different values of the instrumental uncertainty $\Delta \Phi$ have
been chosen.
Values of $\Delta \Phi$ larger than the intra-well separation
$\Delta L$ do not allow to distinguish the two senses of
the superconducting currents: optimal zones of distinguishability
are present only for $\Delta\Phi<\Delta L/2$.
Furthermore, for $\Delta\Phi>\Delta L$, the violations itself disappear.
The plot has been obtained for some values of the instrumental error in a
range of the order of the intra-well distance; moreover, the state has been
prepared with different sequences of initial measurements.
Also, both the filtering functionals (\ref{WVN}) and
(\ref{WA}) have been used.
In all the examined cases, including $\Delta\Phi\ll\Delta L$,
the results are qualitatively similar to the example shown in Fig.~4, as
we will describe in detail in a future paper.

Our result, although obtained for a particular Bell inequality,
should hold in general.
Violations of temporal Bell inequalities stem from a subtle interplay
between the request for resolving the two wells, to assign in an unambiguous
way the sense of the superconducting current of the rf-SQUID,
and the stringent demand for not destroying the coherence of the state
during consecutive measurements
which is at the basis of the superposition principle.
Indeed the  linearity of the quantum
formalism permit superpositions of macroscopically distinct states which
originates the difference from the realistic behaviour.
Any reasonable quantum theory of measurement must introduce
nonunitarity in the time evolution of a repeatedly observed system,
destroying the abovementioned contradiction, as well illustrated
by Feynman in the case of the two-slit experiment.
Therefore violations to Bell inequalities
are not observed either when no measurement is performed
($\Delta \Phi=\infty$) or when the measurement is too strong
($\Delta \Phi\rightarrow 0$). An intermediate regime exists in
which violation of Bell inequalities is possible.
Unfortunately even in this intermediate regime the
violations are not centered, as already remarked in
\cite{LEGG1}, around time intervals between consecutive
measurements equal to multiple of the tunneling period.
On the other hand, as discussed in detail in \cite{CALARCO},
the measurements are quantum nondemolition only for a
periodicity equal to the tunneling period regardless of the
particular shape of the bistable potential.
With demolitive measurements instead, the back-action of the
previous measurements has to be taken into account (as we have done
by introducing the effective uncertainty $\Delta \Phi_{eff}\ge \Delta \Phi$)
ruling out the distinguishability of the
two superconducting current  senses.
The Heisenberg principle, at the heart of quantum theory and based
on classical considerations too, seems to protect Nature from observing
contradictions between it and realism at the macroscopic level.
As a consequence, even if in principle violations of
temporal Bell-like inequalities are
observable, they seem condemned to remain unobserved.
This also requires a revision of the experiments aimed at testing
temporal Bell inequalities proposed \cite{TE} and in preparation.


We acknowledge stimulating communications with G.C. Ghirardi, fundamental
numerical help and a critical reading of the manuscript from C. Presilla.
This work was supported by INFN, Italy.
%
%

%
%
\begin{figure}
\caption{\label{fig1}
Scheme of the simulated sequences of measurements for the calculation
of the correlation probabilities in (\protect\ref{DELTAP}). After a
preparatory sequence of $N=16$
measurements with the optimal periodicity $T=2\pi\hbar/(E_2-E_1)$ and constant
results $\Phi_n\equiv-\Phi_{min}$ (such that $\Delta\Phi_{\it eff}$ has
reached its asymptotic value, as stated in \protect\cite{MEOP2}),
three different series of measurements are performed.
Circles indicate that a measurement takes
place with result of magnitude $\Phi_{min}$ and the sign written within
the circle. Doubled circles indicate the times at
which $\Delta\Phi_{\it eff}$ is calculated.}
\end{figure}

\begin{figure}
\caption{\label{fig2}
Violation parameter $\Delta P$
for the temporal ({\em top}) and spatial ({\em bottom})
Bell inequality. The latter is the already-known analytical result:
$\Delta P(\theta, \phi)= \sin^2(\frac{\theta}{2})-
\cos^2(\frac{\phi}{2})-\cos^2(\frac{\theta+\phi}{2})$, whereas the former is
our numerical result.
A detailed analysis shows that, besides the smaller entity of the
violations, in the first case the regions of violations have an
asymmetrical shape in the $(t_{ab}, t_{bc})$ plane, as a consequence of the
dependence among subsequent measurements (see Fig.~4 for details).
It has been chosen the instrumental error $\Delta \Phi=2<\Phi_{min}$.}
\end{figure}

\begin{figure}
\caption{\label{fig3}
Effective magnetic flux uncertainties $\Delta\Phi_{+-}^{bc}$,
$\Delta\Phi_{++}^{ab}$, $\Delta\Phi_{--}^{ac}$, versus the measurement
times $t_{ab}$ and $t_{bc}$ for each of the three sequences of measurements
schematized in Fig.~1. On top of each graph are superimposed contour plots
of the optimal regions in which the two half-wells are distinguishable,
{\em i.e.} the effective uncertainty is less than the intra-well distance $
\Delta L$. These form periodic parallel bands with different directions in
each case.}
\end{figure}

\begin{figure}
\caption{\label{fig4}
Comparison between the regions of violation of the inequality
(\protect\ref{BELLINEQ}) [shaded areas]
and those in which, for all the three sequences of Fig.~\protect\ref{fig1},
the two half-wells remain distinguishable [small quasi-triangular zones].
The curves are evaluated for three different values of the
instrumental uncertainty ($\Delta \Phi=1,2,4$ as indicated).
Heisenberg islands disappear for $\Delta\Phi \ge 4$; in all the other cases
they have no intersection with the Bell islands.}
\end{figure}

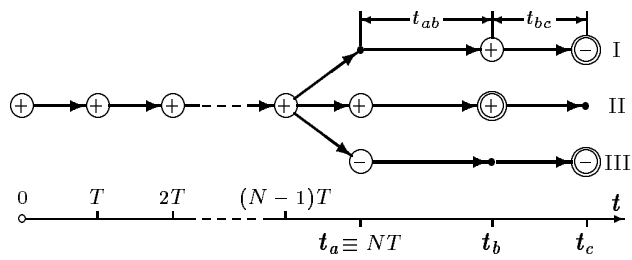
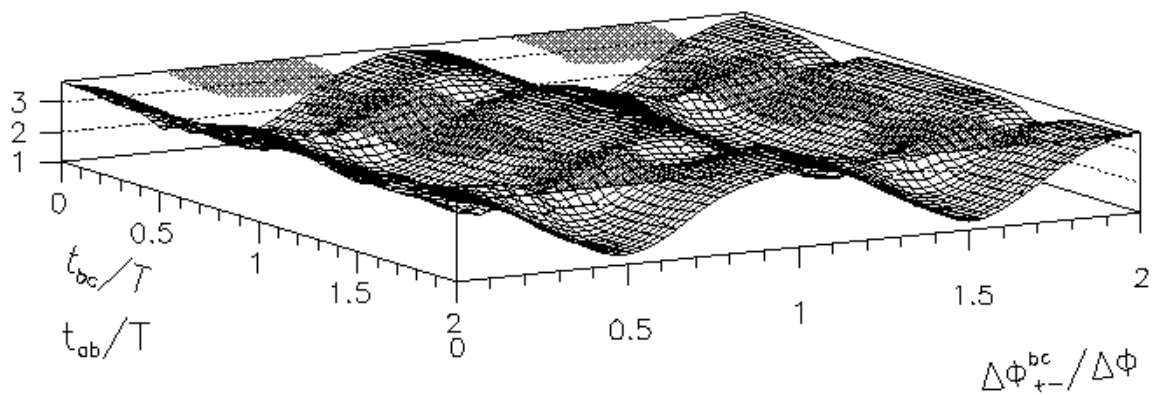
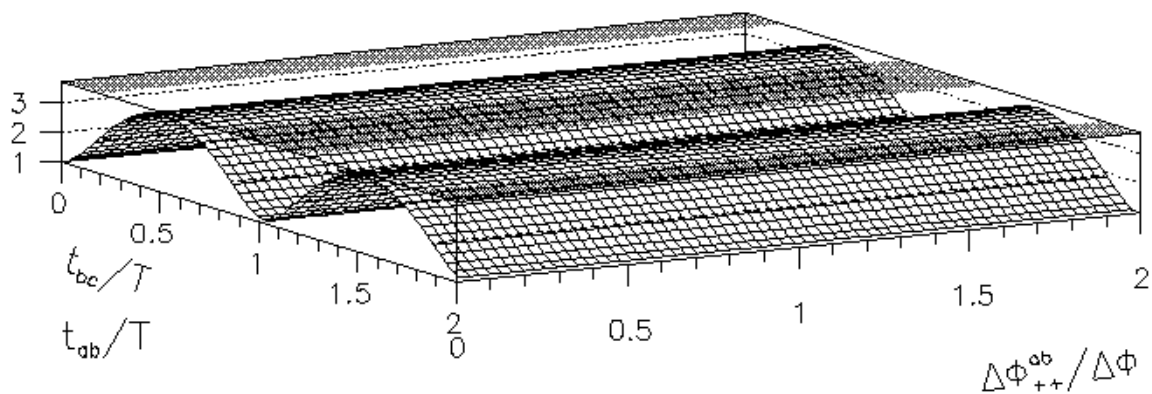
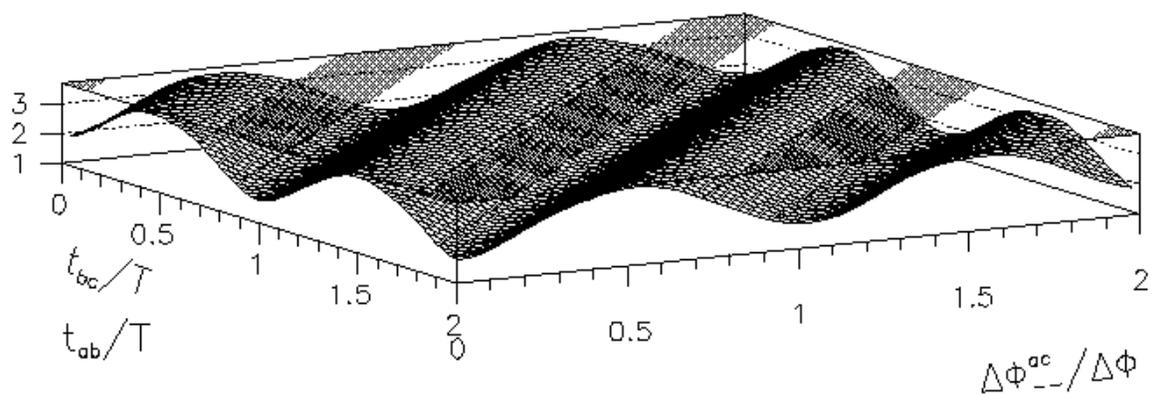
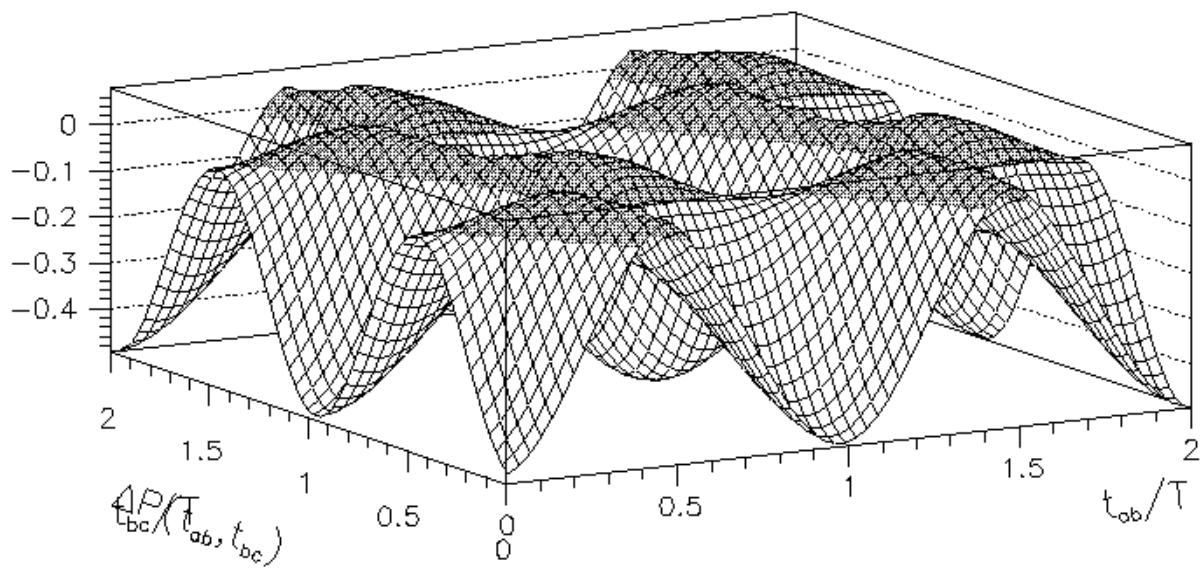
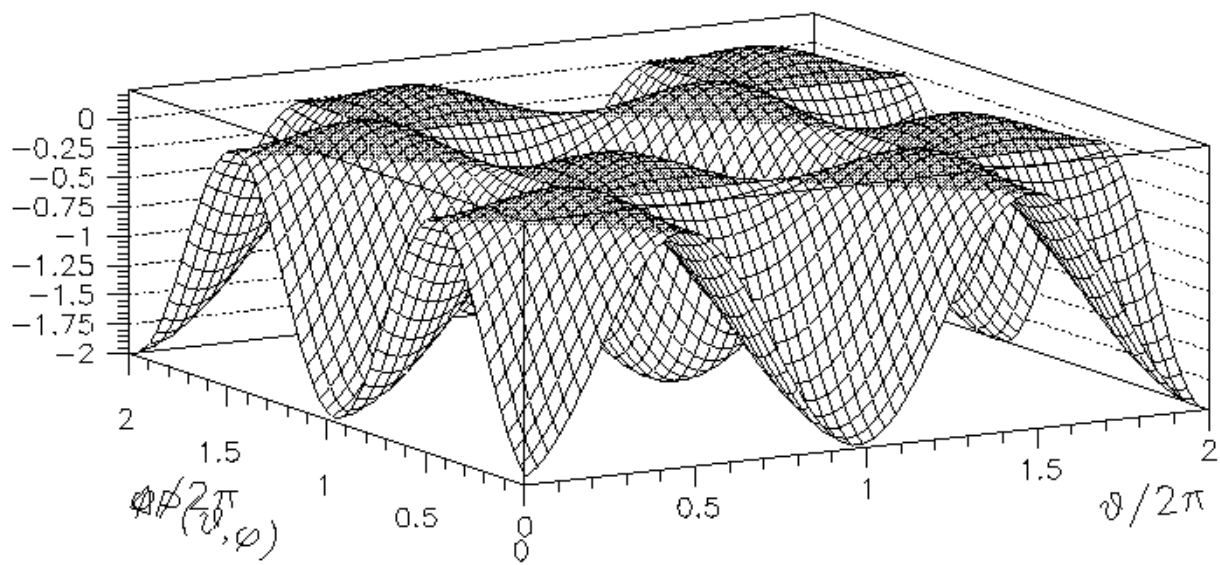
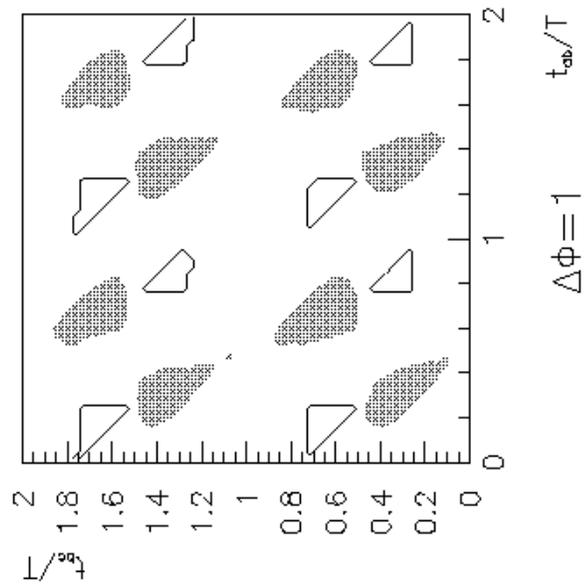
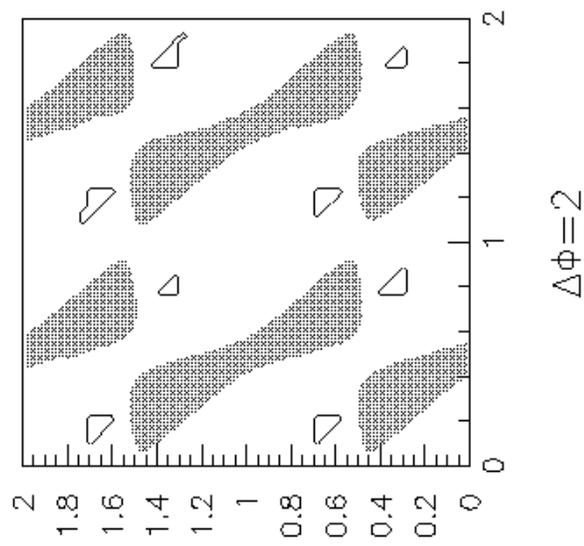
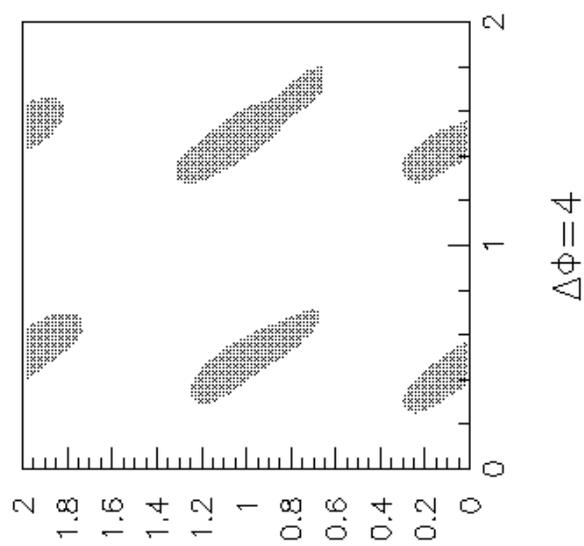
\end{document}